\documentclass[aps,pra,floats,floatfix,twocolumn,superscriptaddress]{revtex4-2}

\usepackage{latexsym}
\usepackage{amsmath,amssymb,amsthm,thmtools}
\usepackage{mathtools,mathrsfs}
\usepackage{amsbsy}
\usepackage{soul}
\usepackage[caption=false]{subfig}
\usepackage{bbold}

\usepackage[utf8]{inputenc}

\usepackage[pdftex]{graphicx}
\graphicspath{{./figures/}}

\usepackage[usenames,dvipsnames,table]{xcolor}
\usepackage{float}

\usepackage[colorlinks=true]{hyperref}
\usepackage[nameinlink]{cleveref}
\hypersetup{
  colorlinks   = true, 
  urlcolor     = green!80!black, 
  linkcolor    = blue, 
  citecolor    = red!80!black 
}

\usepackage{graphicx}

\usepackage[normalem]{ulem}

\usepackage{microtype}

\usepackage{physics}
\usepackage{easyReview}

\newcommand{\parTitle}[1]{\noindent{\color{Mahogany}(\emph{#1})}}
 \renewcommand{\parTitle}[1]{}

\begin{document}

\title{
High fidelity simulation of systems with memory via trajectory mixing under quantum hardware noise
}

\author{Diana A.~Chisholm}
\affiliation{School of Physics, University College Dublin, Belfield Dublin 4, Ireland}
\affiliation{Centre for Quantum Engineering, Science, and Technology, University College Dublin, Dublin 4, Ireland}

\begin{abstract}

Simulating open quantum systems on digital quantum computers typically relies on the use of auxiliary qubits, introducing noisy multi-qubit gates, that severely limit execution on near-term hardware.
In this work, we explore the simulation of non-Markovian dynamics as well as memory channels leveraging the method of trajectory mixing, valid for mixed unitary channels. This allows to obtain higher fidelity for the resulting states, trading entangling gates for a statistical mixture of independent, pure state trajectories.
Using a realistic noise model calibrated to modern quantum processors, we show that trajectory mixing yields higher state fidelity and better preservation of quantum correlations compared to auxiliary qubit implementations.
We further extend this framework to non-Markovian dynamics driven by stochastic Hamiltonians.
This work establishes a practical approach for noise-resilient simulation of open quantum systems with memory.
\end{abstract}

\maketitle

\section{Introduction}


Non unitary evolutions~\cite{breuer_theory_2007, delgado2025quantum} are ubiquitous in quantum science. They emerge naturally whenever the system cannot be adequately isolated from its environment, which occurs commonly in fields such as quantum communication and quantum computing.
It is therefore paramount to be able to efficiently simulate open evolutions in platforms such as quantum computers and quantum simulators.

However, non unitary evolutions cannot be simulated naturally on a quantum computer, given that quantum circuits can only represent unitary evolutions. This makes it necessary to employ specific methods to simulate them.

One of the most widely used methods to do so is the Stinespring dilation~\cite{childs_efficient_2017, wei_efficient_2018}, which states that any quantum evolution (quantum channel) on a system $\mathcal{S}$ can be seen as a unitary evolution $U$ acting jointly on $\mathcal{S}$ and an auxiliary system $\mathcal{A}$, followed by tracing out (ignoring) $\mathcal{A}$.
In other words, we can always express a quantum channel $\Phi(\cdot)$ acting on $\rho_\mathcal{S}$ as $\Phi(\rho_\mathcal{S})=\Tr_\mathcal{A}\left[U\rho_\mathcal{S}\rho_\mathcal{A}U^\dagger\right]$.  

This method however requires the use of multiple auxiliary qubits that interact with the system qubits, resulting in dense interaction topologies~\cite{memon2024quantum}, meaning that a large number of two-qubit gates needs to be used. The implementation of two qubit gates is significantly more noisy than that of single qubit gates, making the Stinespring dilation prone to compound errors. The fewer auxiliary qubits that are used, the better the simulation will be in terms of resilience to noise~\cite{leymann2020bitter}. 
This has led to the development of several methods that can simulate open quantum systems efficiently~\cite{vom_ende_unitary_2019, schlimgen_quantum_2021, ding_simulating_2024, liu_simulation_2025, brearley_linear_2026}, including recent quantum simulation schemes for non-Markovian environments~\cite{seneviratne2025quantum, dan2025simulating} and effective dissipation descriptions~\cite{li2024toward}. The use of auxiliary qubits is however still necessary in general, even if in smaller number compared to the Stinespring dilation.

A specific class of channels, the mixed unitary channels~\cite{rosgen2008additivity} can be simulated without using any auxiliary qubits, by running multiple circuits and statistically mixing the results. We will refer to this as the mixing of trajectories.
This method was shown to be considerably less noisy compared to the Stinespring dilation on real quantum hardware~\cite{peetz_simulation_2024}.

Here, we showcase the fact that mixed unitary channels can be employed to simulate non-Markovian dynamics, as well as channels with memory, via the mixing of trajectories method, allowing for low noise implementations.
We will first show the implementation of this method for pure dephasing evolutions. We then extend it for non-Markovian evolutions as well as channels with memory, both of which are examples of physically meaningful dynamics that can be implemented with mixed unitary channels.
We will discuss the advantage (or lack thereof) that this method offers compared with simulations with auxiliary qubits.
To compare the noise resulting from the two methods, we will break down all the evolutions in a series of single qubit gates as well as the CNOT gate.
Finally, we consider dynamics generated by statistically driven Hamiltonians.
\section{Mixed unitary channels}

\begin{figure*}
    \centering

\subfloat[]{\includegraphics[trim={3cm 1cm 1cm 2cm},clip, width = .44\textwidth]{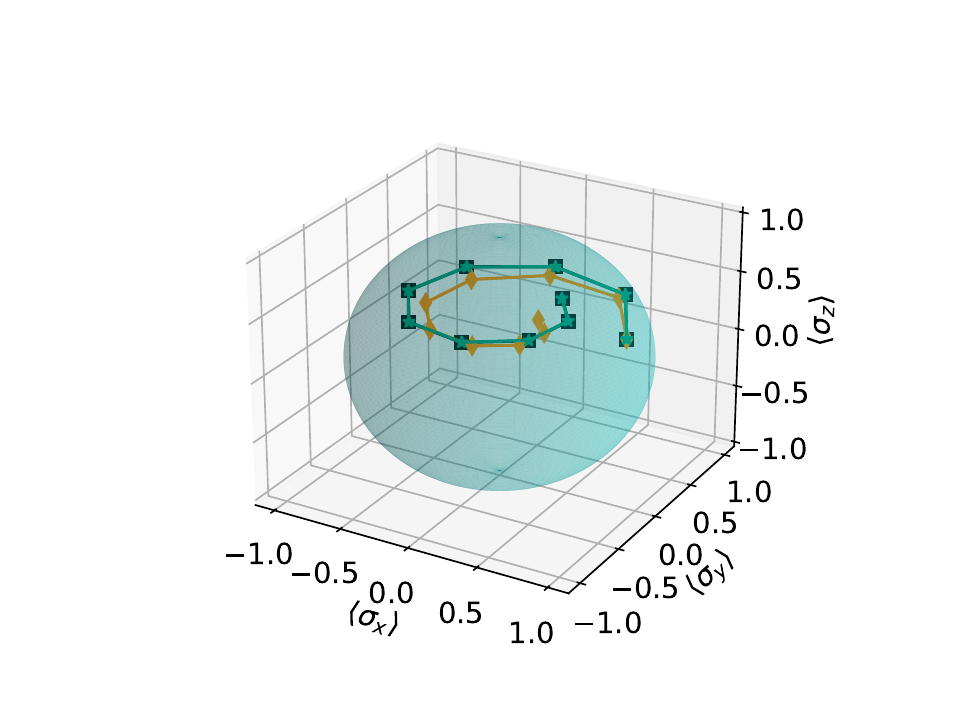}} 
    \subfloat[]{\includegraphics[width=0.55\textwidth]{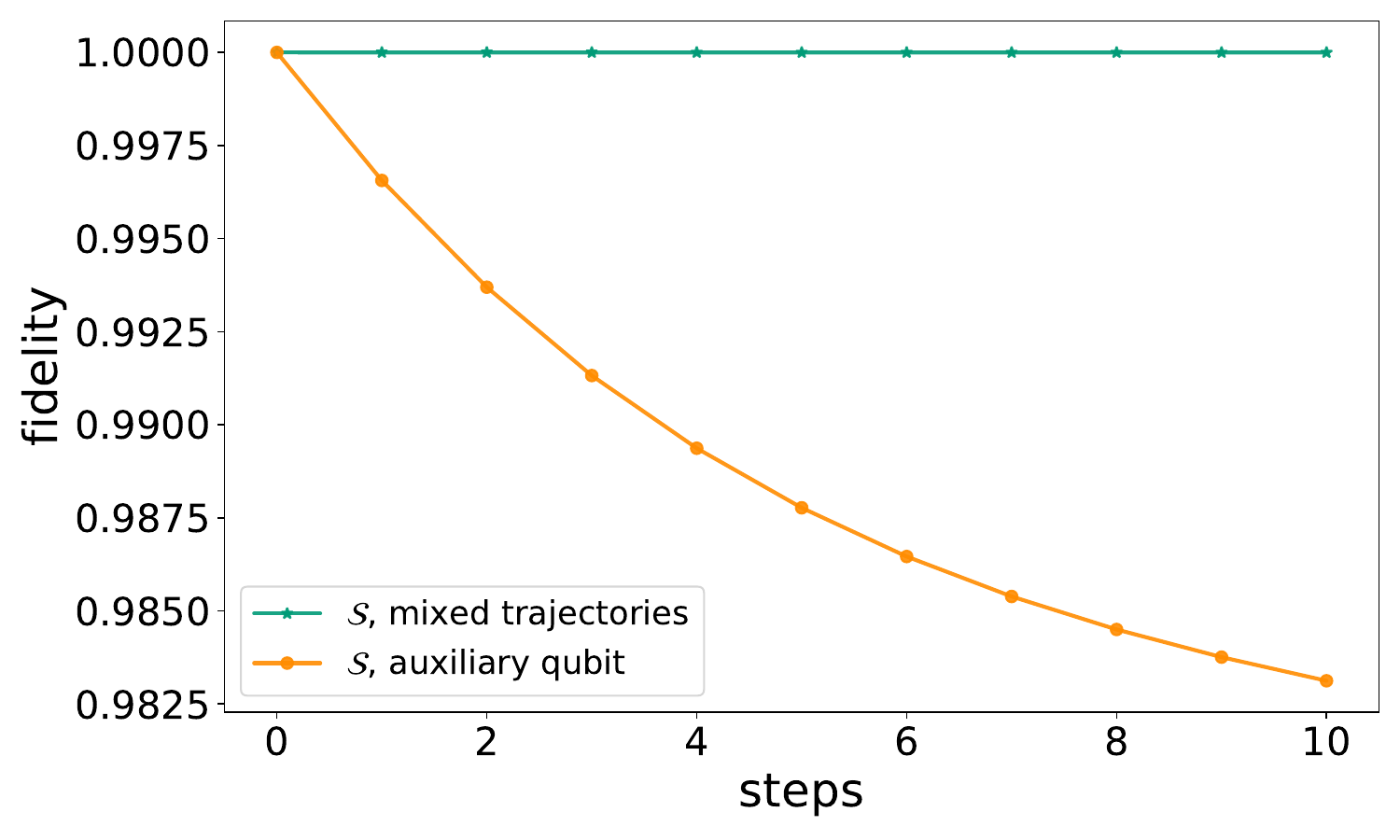}} 
    \caption{\textbf{Simple dephasing}. (a) Dynamics in the bloch sphere for a qubit under pure dephasing, simulated using mixed trajectories (green stars), and using auxiliary qubits (orange diamonds) as well as the noisless dynamics (gray square). Notice how the mixed trajectory dynamics and the noiseless dynamics are almost overlapping with one another, while the auxiliary qubits dynamics is drifting more towards the centre of the sphere. This is further shown in (b), where we show that the dynamics simulated via mixed trajectories (green stars) retains much higher values of state fidelity with the noiseless dynamics, as opposed to the dynamics simulated via auxiliary qubits (orange diamonds).
    }
    \label{fig:dephasing}
\end{figure*}

Suppose we have a channel $\Phi(\cdot)$ in Kraus representation with $K_i$ Kraus operatos $\Phi(\rho)=\sum_i K_i\rho K_i^\dagger$.
The effect of the channel on a state $|\phi\rangle$ can be seen as a statistical mixture of quantum jumps, so that $\Phi\left(|\phi\rangle\langle\phi|\right)=\sum_ip_i|\phi_i\rangle\langle\phi_i|$ where $|\phi_i\rangle=\frac{K_i|\phi\rangle}{\langle\phi|K_i^\dagger K_i|\phi\rangle^{\frac{1}{2}}}$ is the state that ``jumped'' according to the operator $K_i$, and the jump probability is
\begin{equation}
p_{i}=\bra{\phi}K_{i}^{\dagger}K_{i}\ket{\phi}.
\end{equation}
One could therefore simulate the effect of the channel as a collection of pure state trajectories, a rationale behind methods such as the Monte Carlo Wave Function~\cite{dalibard1992wave}.
The issue for an implementation on a real quantum device however, is that the probability depends in general on the state $\ket{\phi}$, and it would then be necessary to evaluate said state every time the channel is applied, in order to determine the jump probabilities $p_i$. This is highly unfeasible in real quantum computers, where states are learned via state tomography, a notoriously expensive task~\cite{james2001measurement, aaronson2007learnability}.

A special case is the mixed unitary channel, where the channel can be written as a statistical mix of unitary evolutions $\Phi(\rho)=\sum_ip_iU_i\rho U^\dagger_i$ and the Kraus operators are therefore rescaled unitary operators $K_i=\sqrt{p_i}U_i$.
In this case, the probability of jumping according to the $i$-th Kraus operator becomes
\begin{equation}
    p_{i}=\bra{\phi}K_{i}^{\dagger}K_{i}\ket{\phi}=\bra{\phi}\sqrt{p_{i}}U_{i}^{\dagger}\sqrt{p_{i}}U_{i}\ket{\phi}=p_{i}
\end{equation}
In other words, the jump probabilities are independent of the state $\ket{\phi}$.

For this reason it is possible to implement the mixed trajectories method only for mixed unitary channels. 
Here, we will consider dynamics that are comprised by a collection of mixed unitary channels applied to the system.
The system evolves according to a unitary evolution, and the same channel is applied to the system at fixed time intervals. This scenario is in the spirit of collision models frequently studied in the context of open quantum systems~\cite{garcia-perez_decoherence_2020, Chisholm_Stochastic_2021, Ciccarello_Quantum_2022, myers2025unifying, fiusa2025queued}.

We stress that mixed unitary channels are only a subclass of all possible quantum channels, and therefore a mixed unitary decomposition is not always possible, a famous example of such a case being the amplitude damping channel.
Nonetheless, many channels of interest within the context of open quantum systems, such as the dephasing and the depolarising channels, are mixed unitary.

The open system dynamics is given by the sum of many pure-states trajectories, one for every combination of $K_i$ Kraus operators.
The total number of possible trajectories is $m^{N}$, where $m$ is the number of Kraus operators that define the mixed unitary channel, and $N$ is the total number of times the mixed unitary channel is applied to the system, and therefore the number of time steps of the dynamics.
For a dephasing channel, this results in $2^{N}$ trajectories. This exponential scaling makes it difficult to simulate long dynamics, but $10$ time steps is within the feasibility of modern quantum computers. We will show later how this restriction can be lifted by performing a stochastic sampling of the trajectories.

We can therefore implement mixed unitary channels without making use of auxiliary qubits, with a significant reduction of the resulting noise.
By considering all possible trajectories, we do not need to assign the probabilities for each jump beforehand. This means that the same collected data can be used to simulate different channel strengths, without the need for further experiments.

We simulate the noise of the gates in the following way: after any single qubit gate we have a probability $p_{T_1}$ that the qubit relaxes (jumps to the $\ket{0}$ state), and a probability $p_{T_2}$ that the qubit loses phases (complete dephasing in the $\sigma_z$ basis). These probabilities are given by $
p_{T_1}=1-e^{-T_G/T_1}$, $p_{T_2}=1-e^{-T_G/T_2}$, with $T_G=60\ ns$ being the gate time, $T1=232\ \mu s$ and $T2=127\ \mu s$.
After the CNOT gate we apply a depolarising channel with probability $p_\mathrm{CNOT}=0.024$.
This noise model is chosen to be comparable to the noise of current quantum processors.

\section{Pure dephasing}

\begin{figure*}
    \centering
    \subfloat[]{\includegraphics[trim={3cm 1cm 1cm 2cm},clip, width=0.44\textwidth]{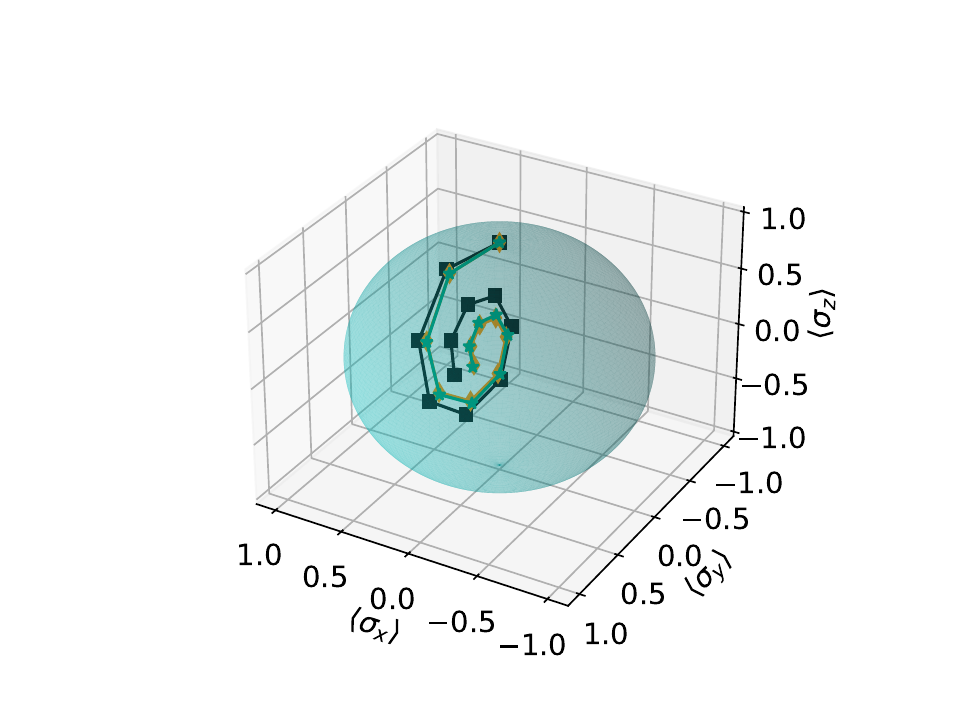}}
    \subfloat[]{\includegraphics[width=0.55\textwidth]{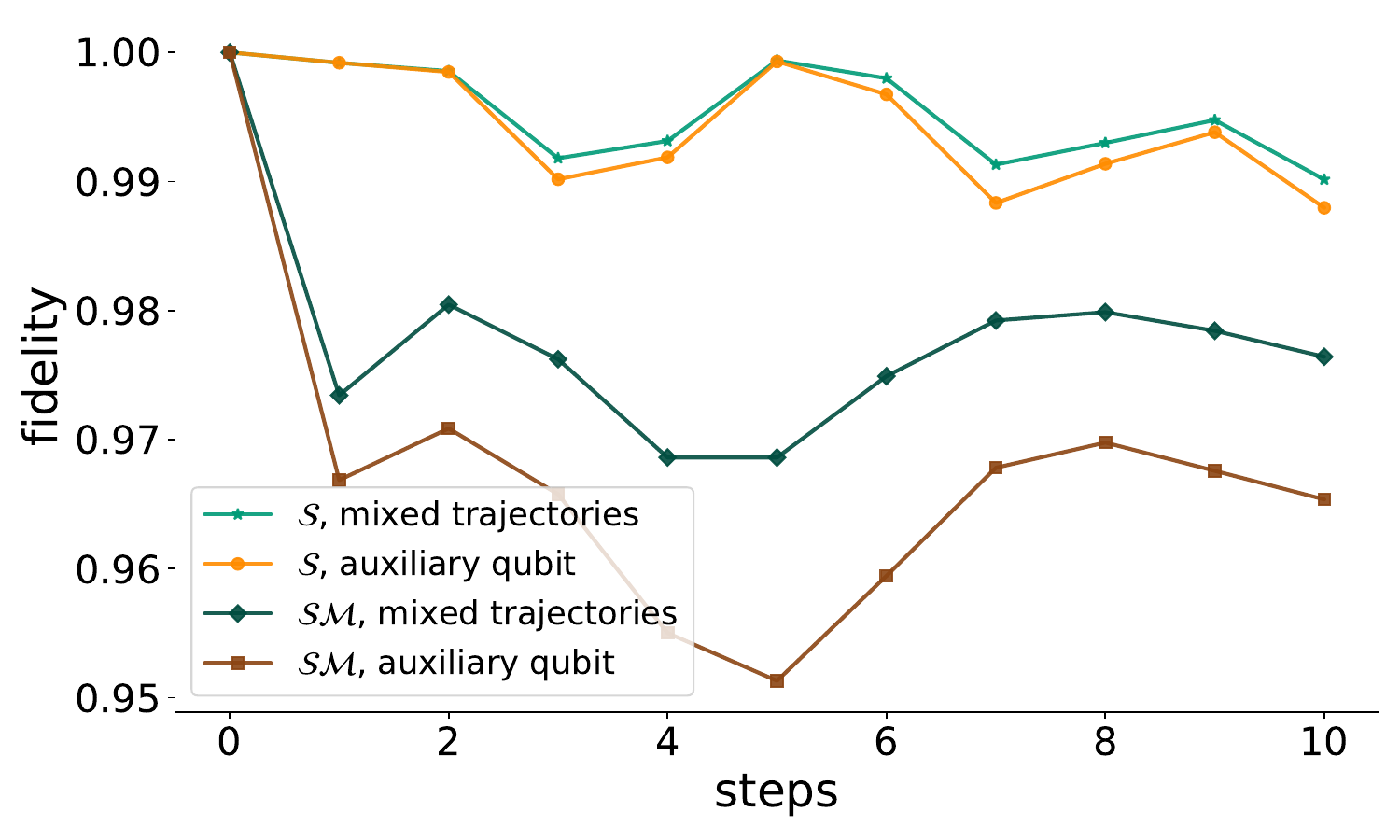}}
    \caption{\textbf{Non-Markovian amplitude damping.} (a) Dynamics in the bloch sphere for a qubit undergoing non-Markovian amplitude damping, simulated using mixed trajectories (green stars), and using auxiliary qubits (orange diamonds) as well as the noisless dynamics (gray square). 
    Here, both the mixed trajectory and the auxiliary qubits dynamics are drifting more towards the centre of the sphere, compared to the noiseless dynamics. In (b) we compare the fidelity, with respect to the noiseless dynamics, of the mixed trajectories and the auxiliary qubits methods. We consider both the system alone (green stars and orange diamonds) and the system and memory together (dark green stars and dark orange diamonds).
    While the mixed trajectories method offers consistently higher fidelity values than the auxiliary qubits method, the difference is not very large when considering the system alone, while it is more noticeable for the system-memory state.}
    \label{fig:nonMarkovian}
\end{figure*}

\begin{figure}
    \centering
    \includegraphics[width=\linewidth]{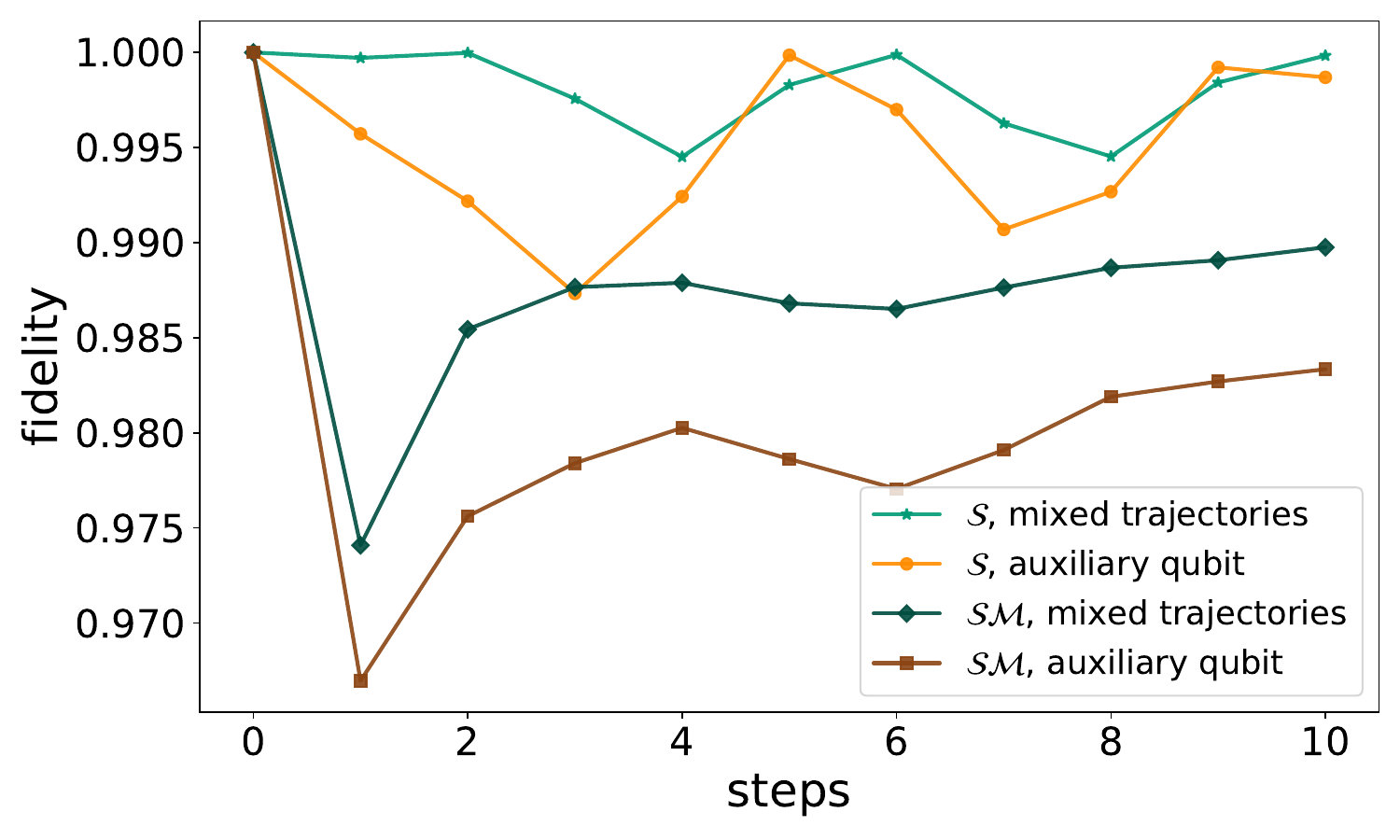}
    \caption{\textbf{Non-Markovian dephasing.} Fidelity values with respect to the noiseless dynamics. Also in this case the mixed trajectories method offers better values of fidelity in general, but the auxiliary qubits method can occasionally perform better while looking at the system only.}
    \label{fig:nonMarkov2}
\end{figure}

As an illustrative example we consider a single qubit undergoing free evolution under the $H=\sigma_z$ Hamiltonian, as well as pure dephasing in the $\sigma_z$ basis.

We choose the initial state of the qubit to be $\sqrt{\frac{7}{10}}|0\rangle+\sqrt{\frac{3}{10}}|1\rangle$.
At each time step, the qubit first evolves with the unitary $U=e^{-i\tau_1\sigma_z}$, and is then subject to the dephasing channel $\Phi(\rho)=p_c\sigma_z\rho\sigma_z+(1-p_c)\rho$.
The free evolution is implemented by applying the $R_z(2\tau_1)$ gate to the system qubit.

For the mixed trajectory method, every time the system is subject to the dephasing channel, we either apply the $\sigma_z$ or the $\mathbb{1}$ operator.
At any given time $t$, for each trajectory $i$ we define $N_U(i, t)$ as the number of times the $\sigma_z$ jump operator has been applied to the system and $N_I(i, t)$ as the number of times the $\mathbb{1}$ operator has been applied. Then the probability for each trajectory is $p_i(t)=p_c^{N_U(i, t)}(1-p_c)^{N_I(i, t)}$. The state of the system at time $t$ is then given by

\begin{equation}
    \rho(t)=\sum_i^{2^N}p_i(t)\rho_i(t).
    \label{Eq3}
\end{equation}
The value of $p_c$, representing the strength of the mixed unitary channel, is only chosen at the stage of mixing the trajectories and plays no role in the simulations, meaning that we can use the same trajectories to simulate different channel strengths.

On the other hand, with the auxiliary qubit method, the channel is implemented by letting the system and the auxiliary qubit (prepared in the $|+\rangle$ state) evolve via the $U=e^{-i\tau_2\sigma_z\sigma_z}$ unitary, with $\tau_2=\arcsin(p_c)$.
This unitary can be implemented as a CNOT gate on the auxiliary qubit controlled by the system, followed by a $R_z(2\tau_2)$ gate on the auxiliary qubit and then another CNOT gate on the auxiliary qubit controlled by the system.

Fig.~\ref{fig:dephasing}~(a) shows 10 steps of the dynamics in the Bloch sphere, simulated using the mixed trajectories method (green stars), the auxiliary qubits method (orange diamonds), as well as the noisless dynamics (gray square). 
We can see how the dynamic simulated by mixing the trajectories closely follows the noiseless dynamics, while the one simulated via auxiliary qubits drifts more toward the centre of the Bloch sphere.
This is further demonstrated by Fig.~\ref{fig:dephasing}~(b) which display the fidelity with respect to the noiseless dynamics of the mixed trajectories method and the auxiliary qubits method. The fidelity of the mixed trajectories method is consistently higher than that of the other one.
Here we have chosen the values of $\tau_1=0.4$ and $\tau_2=0.2$.

\section{Non-Markovian dynamics}

\begin{figure}
    \centering
    \includegraphics[width=\linewidth]{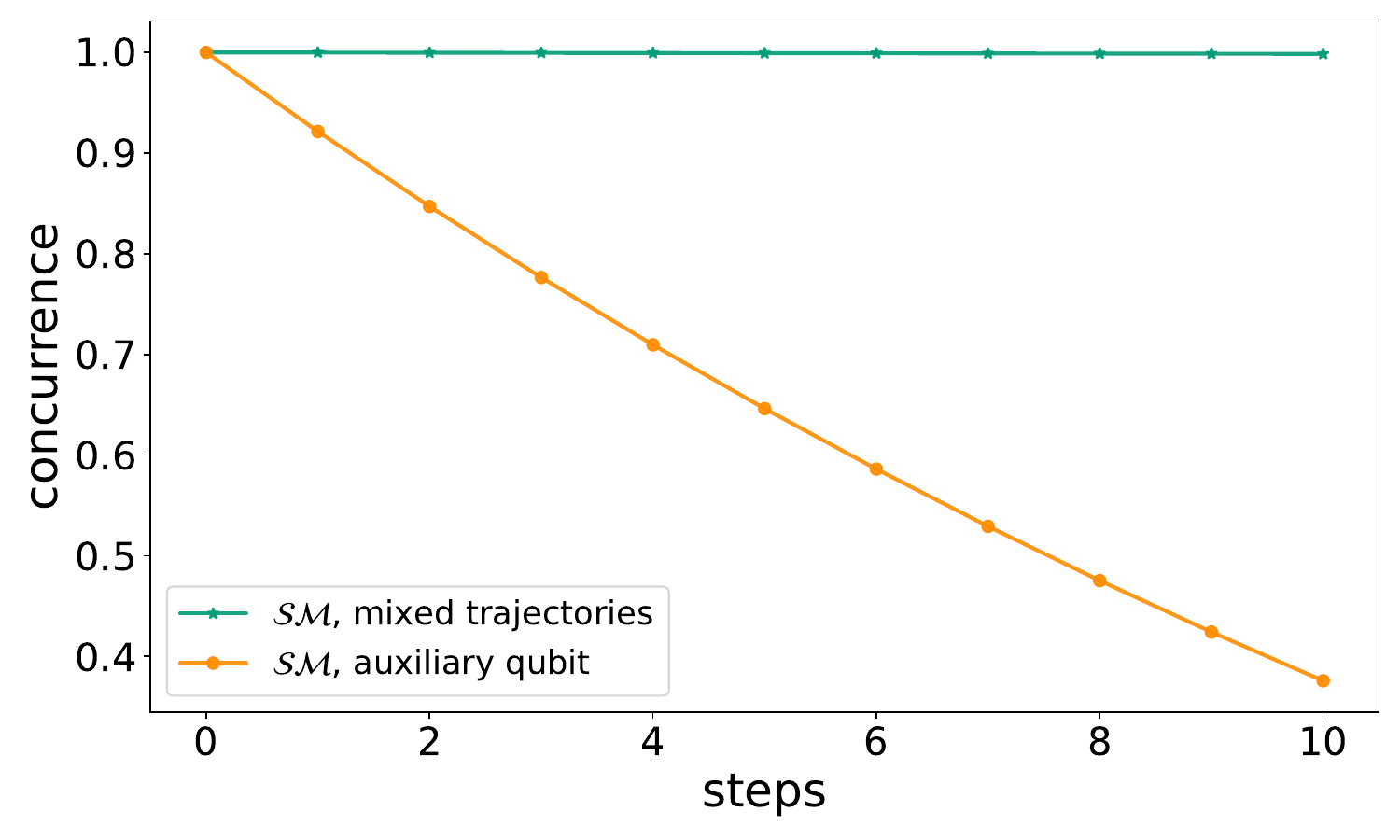}
    \caption{\textbf{Entanglement under memory channels}. We show the concurrence of two qubits initially prepared in a bell state, under the repeated action of a memory channel. When the memory channel is simulated via mixed trajectories entanglement is highly preserved, while it decays considerably when the memory channel is simulated via auxiliary qubits.}
    \label{fig:memory channel}
\end{figure}

As this method allows to simulate any mixed unitary channel, we can also use it to simulate non-Markovian dynamics~\cite{rivas_quantum_2014}.
This can be done by letting the system interact with a memory, such that the system and memory together undergo a Markovian evolution, but once we trace out the memory, the system alone follows a non-Markovian one. This is referred to as a Markovian embedding~\cite{breuer2004genuine, lorenzo2017composite, Campbell_System_2018}.
The introduction of memory qubits is equivalent to reaction coordinates~\cite{strasberg2016nonequilibrium} used in quantum chemistry~\cite{gallina2024simulating,gajewski2026simulating} and open quantum systems, where collective degrees of freedom of the environment are incorporated into the system. While memory qubits describe information backflow, trajectory mixing eliminates the need for additional auxiliary qubits during the dissipative part of the evolution.
In this sense, there is a distinction between auxiliary qubits employed in the Stinespring dilation and memory qubits representing environmental degrees of freedom in a Markovian embedding. Trajectory mixing eliminates the needs for auxiliary qubits, but it does not change the memory size required to capture environmental memory effects. Ultimately, the memory size needed to perform a Markovian embedding remains governed by the environment's complexity~\cite{taranto2020exponential}. Indeed, it is not always possible to perform Markovian embeddings with a finite size memory~\cite{pollock2018non}.
The limitations of the Markovian embedding are entirely indipendent of the way in which the dissipative aspect of the dynamics is simulated.
Moreover, with the trajectory mixing we can only describe non-Markovian scenarios where the dissipative part of the dynamics is described by a mixed unitary channels.

We will show two examples of non-Markovian dynamics. The first is a non-Markovian amplitude damping and the second is a non-Markovian dephasing. 
The non-Markovian aspect of these dynamics can be confirmed by the BLP measure~\cite{breuer_measure_2009} which is positive in both cases.

In the first case, the system and the memory are both initially in the $|0\rangle$ state.
At each time step, the system first evolves with a free evolution via the $R_y(2\tau_1)$ gate, then the system interacts with the memory via a $U=e^{-i\tau_2\sigma_x\sigma_x}$ unitary evolution. This is implemented by a Hadamard gate on the system and on the memory, a CNOT on the memory controlled by the system, a $R_z(2\tau_2)$ on the memory, and then a CNOT on the memory controlled by the system. Finally, the memory undergoes the mixed unitary channel $\Phi(\rho)=p_c\rho+(1-p_c)\sigma_y\rho\sigma_y$, which represents a dephasing in the $\sigma_y$ basis.

The channel is implemented in the same way as in the previous section, by mixing different trajectories in which, each time the channel is applied, we let the memory evolve either with the $\sigma_y$ or with the $\mathbb{1}$ operators.

In the auxiliary qubit method, the memory interacts with the auxiliary qubit, prepared in the $\sqrt{1-p_c}|0\rangle+\sqrt{p_c}|1\rangle$ state, via a Controlled Y gate. This is implemented as an S gate on the memory, followed by a CNOT on the memory controlled by the auxiliary qubit, and then an S gate on the memory.

We show in Fig.\ref{fig:nonMarkovian}~(a) the dynamics in the Bloch sphere, where we can see that the dynamics simulated by mixing the trajectories and the one simulated via auxiliary qubits closely follow each other and drift more towards the centre of the Bloch sphere compared to the noiseless dynamics. Moreover, Fig.\ref{fig:nonMarkovian}~(b) shows that the fidelity for the mixed trajectories method is only marginally higher than the one for the auxiliary qubits method when considering the system, while it offers better improvements when considering the system-memory state. Here we have chosen the values of 
$\tau_1=0.4$, $\tau_2=0.4$ and $p_c=0.1$.

In the second case the system and the memory are both prepared in the $|+\rangle$ state, at each time step they interact with one another via a $U=e^{-i\tau_2\sigma_z\sigma_z}$ unitary, after which the system is subject to the mixed unitary channel $\Phi(\rho)=p_c\rho+(1-p_c)\sigma_y\rho\sigma_y$.

We see in Fig.\ref{fig:nonMarkov2} that the fidelity values for the non-Markovian dephasing follow a similar pattern than the non-Markovian amplitude damping. Here we have chosen $\tau_2=0.4$ and $p_c=0.1$.
For both of the non-Markovian dynamics under consideration, the fidelity can increase in time because the steady state of the dynamics is the maximally mixed state.

\begin{figure*}
    \subfloat[]{\includegraphics[width=0.49\linewidth]{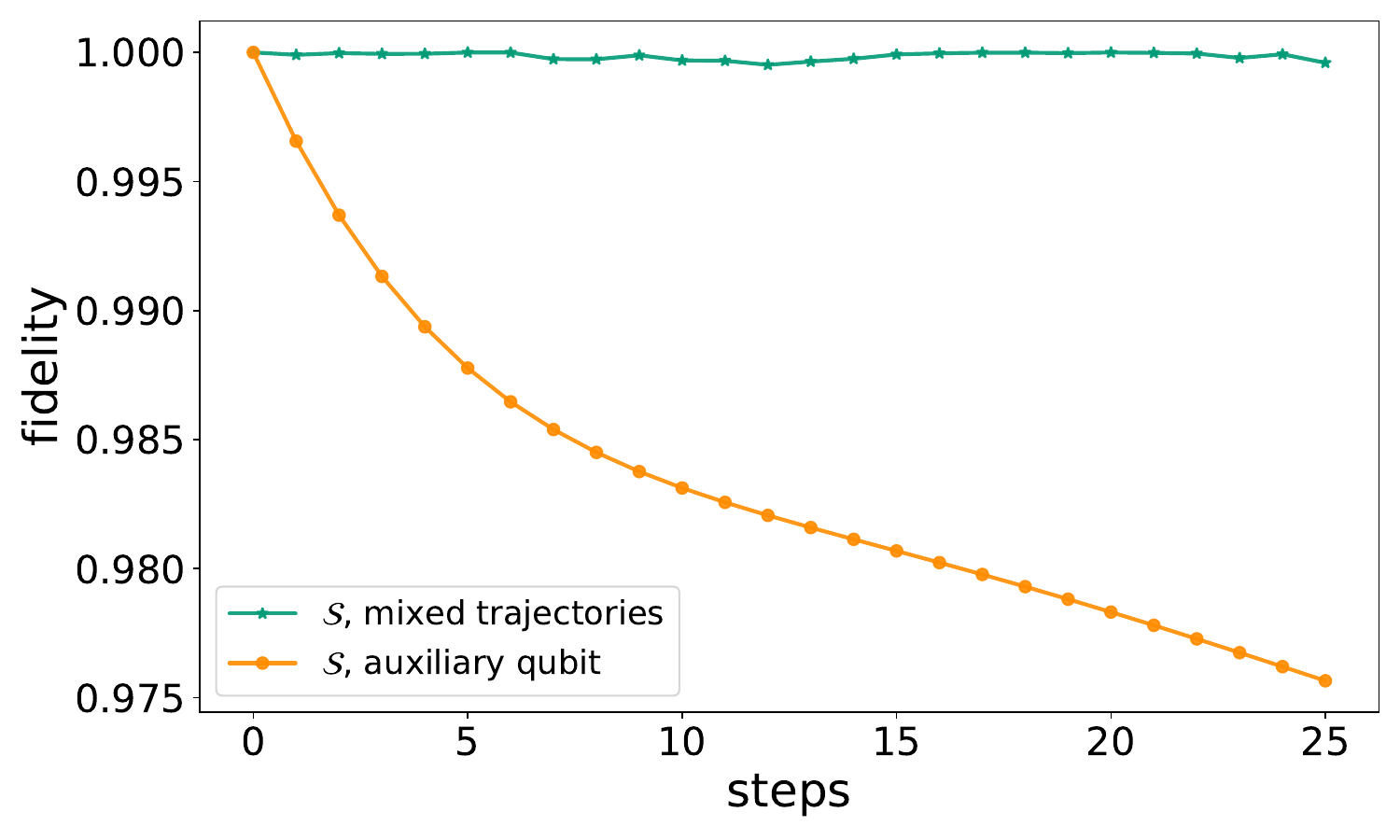}}
    \subfloat[]{\includegraphics[width=0.49\linewidth]{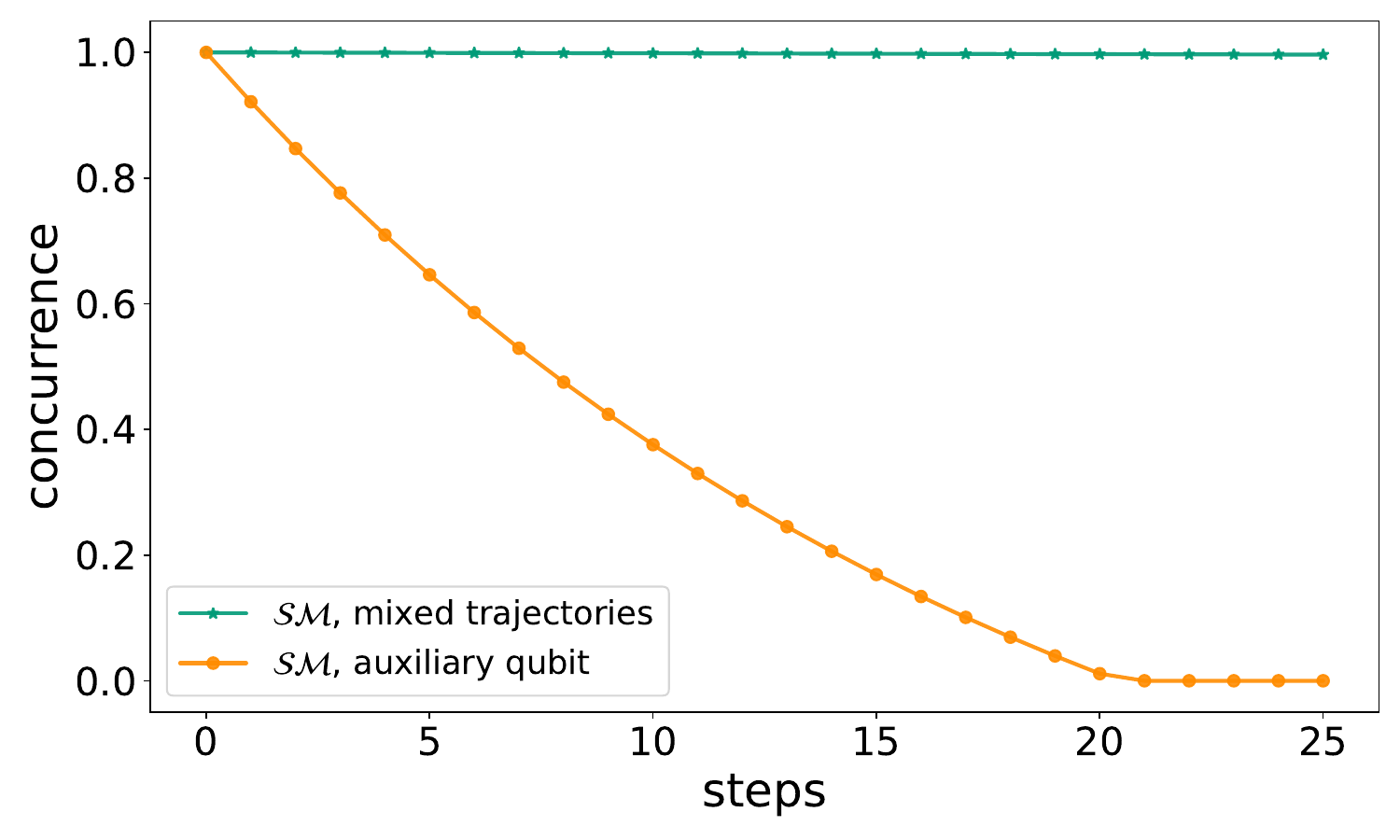}}
    \caption{\textbf{Long time evolutions.} Fidelity with respect to the noiseless dynamics for the simple dephasing (a) and concurrence under repeated action of a memory channel (b), simulated for 25 time steps. Even though we only simulate a fraction of all the possible trajectories, the mixed trajectories method performs significantly better than the auxiliary qubits one.}
    \label{fig:long dynamics}
\end{figure*}
We can see that in these cases, the advantage of the mixed trajectories method is less obvious. While it still offers better fidelity with respect to the exact evolution compared to the use of auxiliary qubits (especially when considering the system-memory joint state), this advantage may not be convenient enough to justify the intrinsic costs of the method.

In the first case, the mixed unitary channel is applied to the memory, while in the second case it is applied to the system. This however does not seem to make much of a difference in terms of fidelity of the dynamics. The fact that system and memory are constantly interacting with each other means that they are never fully protected from the noise of 2-qubit gates.
This suggests that the mixed trajectory may be less useful when simulating environmental interactions in complex (many body) dynamics, which would be noisy even if the channel simulation is efficient.
Non-Markovian evolutions in particular may be ill suited for such methods, considering that the information backflow makes it impossible to protect information locally. Indeed, the mixed trajectories method outperforms the auxiliary qubits method when considering the system and memory together, which follow a Markovian dynamics, rather than the system alone, which follows a non-Markovian one.

We should however also consider that real quantum computers have specific topologies and not all qubits can interact directly with each other. To implement the collision model dynamics we are showing here, one would need to move around the auxiliary quantum states via the use of SWAP gates (which are implemented using three consecutive CNOT gates). Alternatively, qubits can be reset, limiting the need to employ several auxiliary qubits and therefore the need to move states using SWAP gates. The resetting operation however is particularly slow and leads to significant decoherence.
We didn't include these aspects in our simulations for the sake of conceptual simplicity, meaning that the auxiliary qubit simulations are to be interpreted as a ``best case scenario'', implying either high connectivity in the computer, or efficient resetting.

Therefore, while the mixed trajectory method does not seem to be intrinsically superior in this case, it may still be advantageous in a real quantum devices, where these considerations become relevant.

\section{Memory channels}

Memory channels~\cite{macchiavello2002entanglement, macchiavello2004transition} where introduced to describe scenarios where two qubits are sent in rapid succession via a channel that has some memory. Because of this, the channel applies the same random unitary to the two qubits, resulting in a correlated noise.
We consider the case of a dephasing memory channel acting on a two-qubit state $\rho^{ab}$, with two Kraus operators, one being $\sigma_z^a\sigma_z^b$ and the other being the identity operator $\mathbb{1}$, the memory channel acts as $\Phi(\rho^{ab})= p_c\sigma_z^a\sigma_z^b\rho^{ab}\sigma_z^a\sigma_z^b+(1-p_c)\rho^{ab}$.
One key feature of this channel is that it preserves the entanglement of Bell states, 
we can therefore assess the noise of implementing the channel by looking at the concurrence~\cite{hill1997entanglement} of two qubits initially prepared in a Bell state: under the repeated action of the channel, entanglement should remain constant, and if it decays it is entirely due to the noisiness of the quantum gates.

The implementation of the memory channel using the mixed trajectory method is unaffected by the fact that this is now a two-qubit channel, as it is obtained by mixing the trajectories where, at each channel application, the two qubits evolve either with a $\sigma_z$ operator each or with the $\mathbb{1}$ operator, just as detailed in the previous sections.

The memory channel can also be implemented via an auxiliary qubit, letting the two system qubits each evolve via a Controlled Z gate, controlled by the same auxiliary qubit. The fact that the control is the same ensures that the two gates are correlated, so that the two qubits either both evolve under the $\sigma_z$ operator or via the identity.
The Controlled Z gate is implemented by a Hadamard gate on the system, followed by a CNOT gate on the system controlled by the auxiliary qubit, and then a Hadamard gate on the system.

We show in Fig.~\ref{fig:memory channel} the concurrence between two qubits, initially prepared in the Bell state $\frac{1}{\sqrt{2}}|00\rangle+\frac{1}{\sqrt{2}}|11\rangle$, that repeatedly go through the same dephasing memory channel, with $p_c=0.1$, at each time step. The mixed trajectories simulation maintains the concurrence almost unchanged, while it decays significantly for the simulation with auxiliary qubits.
The advantage in term of noise is particularly high in this case, since we are comparing a simulation using only single qubit gates with one that employs two CNOT gates.

\section{Simulating long-time evolutions}

One of the biggest drawbacks of the method shown here is that the number of possible trajectories scales exponentially with the number of times mixed unitary channels are used in the simulation, making long-time evolutions unfeasible.
However, simulating all trajectories is highly unnecessary, after all, in a Monte Carlo Wave Function simulation, one does not keep track of all possible trajectories but simulates a number of them which is considered statistically sufficient to recover the correct dynamics.
We can do the same thing here, in which case, the jump probabilities need to be decided beforehand, as they are needed to sample the trajectories according the the appropriate distribution. Therefore the same simulation does not automatically include different channel strengths.

While we still retain the advantage of higher fidelity for the individual simulated trajectories, their sum does not formally satisfy Eq.~\ref{Eq3} anymore.
Since we are now doing a statistical sampling of the trajectories, the precision of the reconstructed state is governed by Monte Carlo convergence, where the number of trajectories required to achieve a target precision $\epsilon$ scales as $\mathcal{O}(1/\epsilon^2)$, independent of the system size, though the variance of the estimator may depend on the specific observable and evolution time~\cite{plenio1998quantum}.
Meanwhile, the memory requirement remains strictly governed by the complexity of the non-Markovian process, but regardless of the structure of the memory, no additional ancillary qubits are needed to execute the dissipative channel.

We show in Fig.~\ref{fig:long dynamics} the same simulations for the pure dephasing case as well as for the memory channel case, using the same parameters as before, where this time we make the simulation last 25 time steps. 
For the mixed trajectories method, we only sample 1024 trajectories. For the auxiliary qubit method, we still employ an auxiliary qubit for each channel implementation.
Even with this limited sampling, the mixed trajectories approach maintains high fidelity compared to the noiseless evolution, while the fidelity decays constantly for the auxiliary qubits method. 

\section{Stochastically driven Hamiltonian}

\begin{figure}
    \centering
    \includegraphics[width=\linewidth]{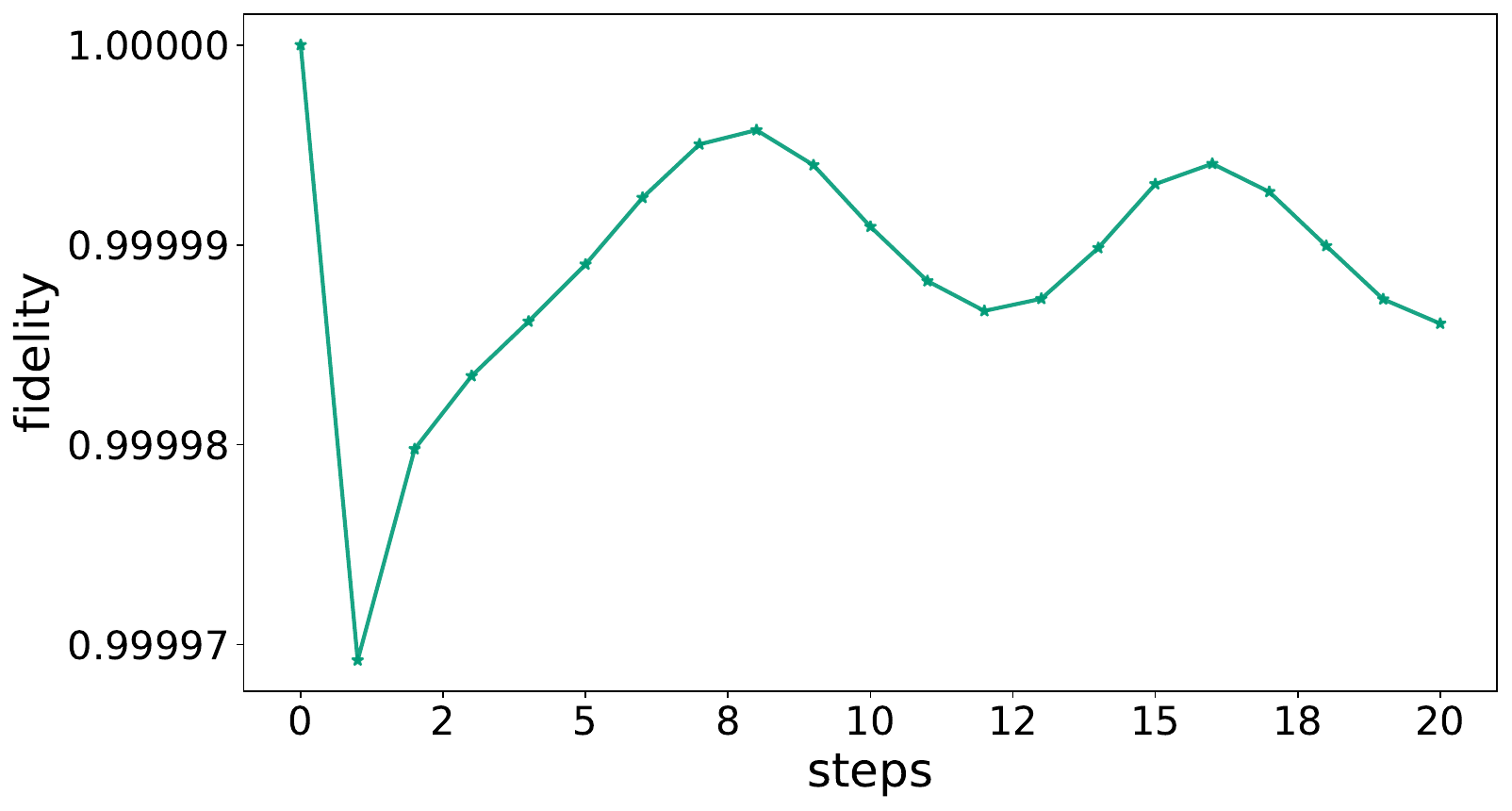}
    \caption{\textbf{Non-Markovian dynamics driven by a stochastic Hamiltonian}. By letting the system evolve under a Hamiltonian that follows a RTN process, we can simulate non-Markovian dynamics without the need of a memory. This results in extremely high fidelities compared to the noiseless dynamics.}
    \label{fig:Real RTN}
\end{figure}

There is another way to directly implement this method to simulate non-Markovian dynamics, and that is to simulate evolutions where the system is driven by a Hamiltonian that follows a stochastic process. We show here the case where a single qubit evolves via the Hamiltonian $H(t)=H_0+V(t)$ where $V(t)$ fluctuates according to a random telegraph noise (RTN)~\cite{paladino20141}. 
The RTN is a stochastic process that jumps at random times between two discrete values.
A process driven by the RTN is subject to a colored bath, and it can therefore exhibit non-Markovianity~\cite{benedetti2016non}.

We have chosen $H_0=0.1\sigma_z$ and $V(t)=0.2X(t)\sigma_z$, where $X(t)$ is the random process that fluctuates between the values of $0$ and $2$, and maintains each value for an average time of $7$ time units. The initial state is $\rho_0=\ket{+}\bra{+}$.

For each iteration of the stochastic process, we can break down the dynamics in time intervals in which $V(t)$ is constant. The time evolution in that time interval is therefore $U=e^{\tau H_0+\tau V(t)}$. Because $H_0$ and $V(t)$ generally do not commute, this unitary cannot be immediately implemented in a quantum computer.
We therefore do a first order trotterisation, where we approximate $e^{\tau H_0+\tau V(t)}$ with $e^{\tau H_0}e^{\tau V(t)}$.

Also in this case the non-Markovian character of the dynamics is proven via the BLP measure.
We are therefore able to simulate non-Markovian dynamics by mixing different trajectories, without employing auxiliary qubits, resulting in dynamics subject to low noise. This is illustrated by Fig.~\ref{fig:Real RTN}, where we show the fidelity between the simulated dynamics and the noiseless dynamics. 

Furthermore, it is possible to connect dynamics driven by classical stochastic processes with dynamics resulting from collision models~\cite{saira2007equivalent, lorenzo2017composite}.

\section{Conclusions}

Here, we have shown that a classical mixing of trajectories can be effectively leveraged to simulate classes of non-Markovian dynamics, memory channels, and stochastically driven Hamiltonians. By trading multi-qubit entangling gates for a statistical ensemble of single-qubit circuits, this approach avoids the noise overhead and qubit routing bottlenecks typical of simulations assisted by auxiliary qubits.
As near-term quantum processors remain limited by multi-qubit gates errors, balancing between circuit noise and number of simulated circuits provides a practical approach for open-system simulations.

\section{Acknowledgements}

We are thankful to Massimo Palma, Mauro Paternostro, Luca Innocenti and Steve Campbell for their support and fruitful discussions. We acknowledge support from Taighde Éireann Research Ireland under Grant No. GOIPD/2025/1353 and from the John Templeton Foundation Grant ID 63626.

\bibliography{Ref}
\end{document}